\documentclass[twocolumn,showkeys,showpacs,preprintnumbers,prd,superscriptaddress,nofootinbib]{revtex4-1}
\usepackage{graphicx}
\usepackage{epsf}
\usepackage{bm}
\usepackage{amsmath}
\usepackage{amsfonts}
\usepackage{amssymb}
\usepackage{epstopdf}
\usepackage{natbib}
\usepackage{color}
\usepackage{verbatim}
\usepackage{multirow}
\usepackage{bm}

\usepackage[colorlinks = true,
            linkcolor = blue,
            urlcolor  = blue,
            citecolor = blue,
            anchorcolor = blue]{hyperref}

\makeatletter\let\expandableinput\@@input\makeatother

\begin{document}
\title{New cosmological constraints on $f(T)$ gravity in light\\
of full Planck-CMB and type Ia supernovae data}

\author{Suresh Kumar}
\email{suresh.math@igu.ac.in}
\affiliation{Department of Mathematics, Indira Gandhi University, Meerpur, Haryana 122502, India}

\author{Rafael C. Nunes}
\email{rafadcnunes@gmail.com}
\affiliation{Instituto de F\'{i}sica, Universidade Federal do Rio Grande do Sul, 91501-970 Porto Alegre RS, Brazil}
\affiliation{Divis\~ao de Astrof\'isica, Instituto Nacional de Pesquisas Espaciais, Avenida dos Astronautas 1758, S\~ao Jos\'e dos Campos, 12227-010, SP, Brazil}

\author{Priya Yadav}
\email{priya.math.rs@igu.ac.in}
\affiliation{Department of Mathematics, Indira Gandhi University, Meerpur, Haryana 122502, India}

\begin{abstract}
We investigate two new observational perspectives in the context of torsional gravitational modification of general relativity, i.e., the $f(T)$ gravity: i) We use Pantheon data of type Ia supernovae  motivated by a time variation of the Newton’s constant on the supernovae distance modulus relation, and find that a joint analysis with Baryon Acoustic Oscillations (BAO) and Big Bang Nucleosynthesis (BBN), i.e., Pantheon+BAO+BBN, provides constraints on the effective free parameter of the theory to be  well compatible with the $\Lambda$CDM prediction; ii) We present the framework of $f(T)$ gravity at the level of linear perturbations with the phenomenological functions, namely the effective gravitational coupling $\mu$ and the light deflection parameter $\Sigma$, which are commonly used to parameterize possible modifications of the Poisson equation relating the matter density contrast to the lensing and the Newtonian potentials, respectively. We use the available Cosmic Microwave Background (CMB) data sets from the Planck 2018 release to constrain the free parameters of the $f(T)$ power-law gravity and $\Lambda$CDM models. We find that CMB data, and its joint analyses with Pantheon and BAO data constrain the $f(T)$ power-law gravity scenario to be practically indistinguishable from the $\Lambda$CDM model.  We obtain the strongest limits ever reported on $f(T)$ power-law gravity scenario at the cosmological level.

\end{abstract}

\maketitle
\section{Introduction}
\label{sec:intro}

Several extensions of general relativity (GR) have been proposed (see \cite{Pedro_2012,Ishak_2018,Nojiri_2017,Olmo_2021,Louis_2020} for a review) and exhaustively investigated to explain the observational data in cosmology and astrophysics. In particular, the additional gravitational degree(s) of freedom from the modified gravity (MG) models quantify extensions of the $\Lambda$CDM cosmology, and can drive the accelerated expansion of the Universe at late times. Several of these extensions have been found to fit the data well, and thereby leading to a possible theoretical degeneracy among several proposals. Among viable candidates for MG theories,  modifications starting from the torsion-based formulation, and specifically from the Teleparallel Equivalent of General Relativity \cite{J. W. Maluf}, have been of great interest in recent years. Since in this theory the Lagrangian is the torsion scalar $T$, the simplest modification is the $f(T)$ gravity (see \cite{EDV_2021,Cai_2016,MK_2019} for a review).

On the other hand, we currently have increasingly accurate measurements of the cosmological parameters that challenge the consensus on the $\Lambda$CDM  model \cite{snowmass_r}. Certainly, the most significant tension with the standard model prevision is the observed value of the present cosmic expansion rate, quantified by the Hubble constant, $H_0$. As well known, assuming the minimal $\Lambda$CDM scenario, Planck-CMB data analysis provides $H_0=67.4 \pm 0.5$ km s$^{-1}$Mpc$^{-1}$ \cite{planck2020}, which is in $\sim$ $5\sigma$ tension with the SH0ES team local measurement $H_0 = 73.30 \pm 1.04$ km s$^{-1}$Mpc$^{-1}$~\cite{Riess_H0}.
Additionally, many other late time measurements are in agreement with a higher value for the Hubble constant (see the discussion in~\cite{Di_Valentino_2021_H0,Perivolaropoulos_H0,snowmass_r}). Motivated by these observational discrepancies, unlikely to disappear completely by introducing multiple and unrelated systematic errors, it has been widely discussed in the literature whether new physics beyond the standard cosmological model can solve the $H_0$ tension (see \cite{Di_Valentino_2021_H0,Perivolaropoulos_H0,snowmass_r,collgain_2021} and references therein for a review). On the other hand, it has been argued that the $H_0$ tension is actually a tension on the supernovae (SN) absolute magnitude $M_B$ \cite{George_2021,David_21}, because the SH0ES $H_0$ measurement comes directly from $M_B$ estimates. The CMB constraint on the sound horizon to the SN absolute magnitude $M_B$ using the inverse distance
ladder predicts $M_{B}= -19.401 \pm 0.027$ mag \cite{VM_20}, while
the SN measurements from SH0ES corresponds to $M_B = - 19.244 \pm 0.037$ mag \cite{David_21}. These measurements are at 3.4$\sigma$ tension. Thus, as argued in \cite{George_2021,David_21,Rafael_21,DC_22} rather than explaining the $H_0$ tension, one should instead focus on the SN absolute magnitude tension, because this is what the Cepheid calibrations are designed to measure. Some attempts to explain the $H_0$ tension by modifying the GR have already been made in the literature (see \cite{Di_Valentino_2021_H0,Perivolaropoulos_H0,snowmass_r} for recent reviews).

MG scenario has an impact on the astrophysics of SN, by allowing the gravitational constant to vary with redshift $z$,  which can induce a redshift dependent effect on the peak luminosity of the SN from the mass of the white dwarf progenitors \cite{LA_99,Bravo_1}. The correction due to the evolution of intrinsic luminosity in the value of Newton’s gravitational constant $G$ has been studied, and it has also been utilised to place constraints on several models \cite{RL_5,RSW_18,HD_20,DS_20}. This means that for accurate cosmological studies of gravity, we must carefully consider modified gravity’s impact on SN Ia astrophysics and its implication in terms of cosmological parameter inference. The first main aim of this work is to make the above mentioned corrections in the $f(T)$ gravity context and study the resulting implications.

It is well known that modification to GR can induce significant changes on the growth of density fluctuations and the large scale structures, as well as on the CMB anisotropies, among several other traces and observations (see \cite{Pedro_2012,Ishak_2018} for a review). These potential deviations are commonly encoded in the phenomenological functions $\Sigma$ and $\mu$ that parametrize modifications of the perturbed Einstein’s equations relating the matter density contrast to the lensing and the Newtonian potential, respectively \cite{Bean_2010,Gong_2009,Hu_2007,Edmund_2008}. The second main aim of this work is to obtain new observational constraints on the $f(T)$ gravity using the full Planck-CMB data, after deriving the relation for the functions $\Sigma$ and $\mu$ in the $f(T)$ gravity framework. The $f(T)$ gravity has been intensively investigated using geometrical data after computing the modified expansion rate of the Universe $H(z)$ \cite{Benetti_2020,Aliya_2022,David_2020,Amr_2021,RCN_2020,Salvatore_2017,Mota_2020,Cai_2020,SB_2018,Said_2020,Zant_2018,Santos_2021,Nashed_2019,Briffa_2022,ZZ_2021,AP_2021,FK,QI}, and also on sub-horizon scales by considering measurements of the growth rate \cite{Nunes_2020,Chen_2020,Fotios_2019,Nelson_2018,Rocco_2018,Sultana_2021}. The $f(T)$ teleparallel gravity is investigated previously with different observational data in \cite{Benetti_2020,Aliya_2022,David_2020,Amr_2021,RCN_2020,Mota_2020,Cai_2020,SB_2018,Zant_2018,Santos_2021,Nashed_2019,Briffa_2022,ZZ_2021,FK,SC,RC,rf_17,QI,Fotios_2019,Bing,FBM_2022}, and in particular with full  CMB data  in \cite{Benetti_2020,Hashim_2021,Rafael_2018}. The new ingredient in the present work is to quantify the MG effects parameterized by phenomenological functions $\Sigma$ and $\mu$, which are intensively used in the literature for others MG models, as well as to observe the effects of new corrections on the SN Ia as described above. With these perspectives, we aim to present the most robust constraints on $f(T)$ gravity using the latest Planck-CMB, BAO, BBN and Pantheon data. 

The manuscript is organized as follows. In section \ref{ft}, we briefly review the $f(T)$ gravity and its cosmology, where the scalar perturbative evolutions in terms of $\Sigma$ and $\mu$ are described along with one of the most-used $f(T)$ gravity power-law model. In section \ref{data}, we describe the data sets used for our analysis and the methodology adopted. In section \ref{results}, we present our main results and disscussion, focusing on the power-law model and the effects on the CMB power spectrum. Finally, in section \ref{final} we summarize our results.

\section{$f(T)$ gravity}
\label{ft}
The action of a generalized teleparallel gravity can be written as
\begin{equation}
    \mathcal{S}=\frac{1}{2\kappa^2}\int d^{4}x\, |e|f(T)+\mathcal{S}_{M},
    \label{action}
\end{equation}
where $T$ denotes the teleparallel torsion scalar; $|e|=\sqrt{-g}=\det\left({e}_\mu{^a}\right)$; $\kappa\equiv \sqrt{8\pi G}$, where $G$ is Newton's constant; and $\mathcal{S}_{M}$ is the action of matter fields. Hereafter, we consider $f(T)= T + F(T)$ and $\mathcal{S}_{M}$ to run over the matter fields, i.e, baryons, dark matter, photons and neutrinos.

We assume that the spatial background geometry of the universe is of a flat Friedmann-Lema\^{\i}tre-Robertson-Walker (FLRW) metric. Hence, we take the Cartesian coordinate system ($t;x,y,z$) and the diagonal vierbein
\begin{equation}
   {e_{\mu}}^{a}=\textmd{diag}\left(1,a(t),a(t),a(t)\right),
\label{tetrad}
\end{equation}
where $a(t)$ is the scale factor of the universe. The above vierbein generates the flat FLRW spacetime metric
\begin{equation}
   ds^2=dt^{2}-a(t)^{2}\delta_{ij} dx^{i} dx^{j}.
\label{FRW-metric}
\end{equation}
This defines the teleparallel torsion scalar as
\begin{equation}
T=-6 H^2. \label{eq:Torsion_sc}
\end{equation}

Variation of the action with respect to the vierbeins provides the field equations as
\begin{eqnarray}
\label{eom}
&&\!\!\!\!\!\!\!\!\!\!\!\!\!\!\!
e^{-1}\partial_{\mu}(ee_A^{\rho}S_{\rho}{}^{\mu\nu})(1+F_{T})
 +
e_A^{\rho}S_{\rho}{}^{\mu\nu}\partial_{\mu}({T})F_{TT}
\nonumber\\
&& -(1+F_{T})e_{A}^{\lambda}T^{\rho}{}_{\mu\lambda}S_{\rho}{}^{\nu\mu}+\frac{1}{4} e_ { A
} ^ {
\nu
}[T+F({T})] \nonumber \\
&&= 4\pi Ge_{A}^{\rho}
\left[{\mathcal{T}^{(m)}}_{\rho}{}^{\nu}+{\mathcal{T}^{(r)}}_{\rho}{}^{\nu}\right],
\end{eqnarray}
with $F_{T}=\partial F/\partial T$, $F_{TT}=\partial^{2} F/\partial T^{2}$,
and where ${\mathcal{T}^{(m)}}_{\rho}{}^{\nu}$ and  ${\mathcal{T}^{(r)}}_{\rho}{}^{\nu}$
are  the matter and radiation energy-momentum tensors respectively.

Inserting the vierbein (\ref{tetrad}) into the field equations
(\ref{eom}), we acquire the Friedmann equations as
\begin{eqnarray}\label{background1}
&&H^2= \frac{8\pi G}{3}(\rho_ {\rm m}+\rho_{\rm r})
-\frac{f}{6}+\frac{TF_T}{3},\\\label{background2}
&&\dot{H}=-\frac{4\pi G(\rho_{\rm m}+P_{\rm m}+\rho_{\rm r}+P_{\rm r})}{1+F_{T}+2TF_{TT}},
\end{eqnarray}
with $H\equiv\dot{a}/a$ the Hubble parameter, and where we use dots to denote derivatives with respect to cosmic time $t$.

Observing the form of the first Friedmann equation (\ref{background1}), we deduce that
in $f(T)$ cosmology  we acquire  an effective dark energy sector of
gravitational origin. In particular, we can define the effective
dark energy density as

\begin{eqnarray}
 \rho_{\rm DE}\equiv\frac{3}{8 \pi G} \left(-\frac{f}{6}+\frac{TF_T}{3} \right).
\end{eqnarray}

In what follows, a subindex zero attached to any quantity implies its value at the present time. In this work, we are interested in confronting the model with observational data. Hence, we firstly define 
\begin{eqnarray}
\label{THdef3}
\frac{H^2(z)}{H^2_{0}}=\frac{T(z)}{T_{0}},
\end{eqnarray}
with $T_0\equiv-6H_{0}^{2}$.

Therefore, using additionally that
$\rho_{\rm m}=\rho_{\rm m0}(1+z)^{3}$, $\rho_{\rm r}=\rho_{\rm r0}(1+z)^{4}$, we
re-write the first Friedmann equation (\ref{background1}) as \cite{SN_2013}
\begin{eqnarray}
\label{Mod1Ez}
\frac{H^2(z,{\bf r})}{H^2_{0}}=\Omega_{\rm m0}(1+z)^3+\Omega_{\rm r0}(1+z)^4+\Omega_{\rm F0} y(z,{\bf r})
\end{eqnarray}
where
\begin{equation}
\label{LL}
\Omega_{\rm F0}=1-\Omega_{\rm m0}-\Omega_{\rm r0} \;,
\end{equation}
with $\Omega_{\rm i0}=\frac{8\pi G \rho_{\rm i0}}{3H_0^2}$, the corresponding
density parameter at present. In this case, the effect of the $f(T)$ modification  is
encoded in the function  $y(z,{\bf r})$ (normalized to
unity at   present time), which depends on $\Omega_{\rm m0},\Omega_{\rm r0}$, and on the
$f(T)$-form parameters $r_1,r_2,...$, namely \cite{SN_2013}:
\begin{equation}
\label{distortparam}
 y(z,{\bf r})=\frac{1}{T_0\Omega_{\rm F0}}\left(f-2Tf_T\right).
\end{equation}

It is interesting to note that the additional corrections on the effective Friedman equation (\ref{Mod1Ez})  is a function of the Hubble function only.

Now we describe how the linear scalar perturbations evolve in the context of $f(T)$ gravity. The most general linear scalar perturbations of the vierbein can be written as \cite{Tomi_2018}  \footnote{We changed the metric potentials $\phi$ $\leftrightarrow$ $\psi$ in order to match with the notations presented in \cite{Ma_1995}, which is usually assumed in the Boltzmann codes.}
\begin{eqnarray}
e^0_0 & = & a(\tau)\cdot\left(1+\psi\right),\\
e^0_i & = & a(\tau)\cdot \partial_i \zeta,\\
e^a_0 & = & a(\tau)\cdot \partial_a \zeta,\\
e^a_j & = & a(\tau)\cdot \left((1-\phi)\delta^a_j+\epsilon_{ajk}\partial_k s\right),
\end{eqnarray}
where $\phi$ and $\psi$ are the two potentials describing the scalar modes of the metric perturbations (fixed to the Newtonian gauge). The function $\zeta$ represents the scalar part of a Lorentz boost, and $s$ is the scalar part of a spatial rotation. The variable $\tau$ stands for conformal time. The complete set of linear perturbations (scalar, vector and tensor modes) are well described in \cite{Ma_1995}. In this present work, we deal with the scalar and tensor modes.

First, the Poisson equation in GR, obtained from the combination of time-time and time-space components of the perturbed Einstein equation, reads

\begin{equation}
\label{Poisson}
- k^2 \phi = 4 \pi G a^2 \sum \rho_i \Delta_i,
\end{equation}
while the anisotropic space-space component yields

\begin{equation}
\label{ij_term}
- k^2(\psi-\phi)  = 12 \pi G a^2 \sum \rho_i (1+w_i) \sigma_i,
\end{equation}
where in the above equations $\Delta_i= \delta_i + 3 \mathcal{H} (1+w_i) \theta_i/k^2 $ is the rest-frame density perturbation of matter species $i$, $\sigma_i$ is the anisotropic shear stress, and $\theta_i$ is the divergence of the peculiar velocity. The function $\mathcal{H}$ is the Hubble function in the conformal time, which is related to cosmic time by $\mathcal{H}= aH$. Hereafter prime means derivative with respect to conformal time. In GR, the lensing equation is written as

\begin{equation}
\label{lensing}
- k^2(\psi+\phi)  = 8 \pi G a^2 \sum \rho_i \Delta_i.
\end{equation}

Now,  we obtain the same equation for the $f(T)$ gravity. Combining the  symmetric part and the mixed components of the perturbation equations described in \cite{Tomi_2018}, we obtain the Poisson equation 

\begin{eqnarray}
\label{poisson}
-k^2 \phi &=& 4 \pi \mu_T G a^2 \sum \rho_i \Delta_i,
\end{eqnarray}
while the spatial part and the off-diagonal components yield

\begin{eqnarray}
\label{lensing}
-k^2\left(\psi-R \phi\right) &=& 12\pi \mu_T G a^2  \sum \rho_i (1+w_i) \sigma_i\nonumber\\ && + 12 \pi \mu_T G \Xi a^2 \mathcal{H} \sum \rho_i (1+w_i) \theta_i.
\end{eqnarray}
In the above equations, we define the following quantities:

\begin{equation}
\mu_T = \frac{1}{f_T},
\end{equation}

\begin{equation}
\Xi = 12 (\mathcal{H}'-\mathcal{H}^2) \mu_T f_{TT},
\end{equation}

\begin{equation}
R = 1+\frac{3 \Xi}{k^2 a^2}\left[\mathcal{H}^2 \Xi + (\mathcal{H}'-\mathcal{H}^2) a^2 \right].
\end{equation}

As expected for $\mu_T=1$ and $\Xi=0$, we recover GR. The lensing equation in $f(T)$ gravity can still be written in the form 

\begin{eqnarray}
-k^2\left(\psi+\phi\right) &=& 8 \pi \mu_T G a^2 S  ,\label{lensing_ft2}
\end{eqnarray}
where
\begin{eqnarray}
S &=& \sum \rho_i \Delta_i \Big[ \frac{(1+R)}{2} + \frac{3}{2}\frac{ \Xi \mathcal{H} \sum \rho_i (1+w_i) \theta_i}{\sum \rho_i \Delta_i} \Big].
\end{eqnarray}
Here we can identify

\begin{eqnarray}
\Sigma_T = \frac{\mu_T}{2} \Big[ (1+R) + \frac{3 \Xi \mathcal{H} \sum \rho_i (1+w_i) \theta_i}{ \sum \rho_i \Delta_i} \Big],
\end{eqnarray}
to define the lensing equation in its usual form

\begin{eqnarray}
\label{lensing_equation}
-k^2\left(\psi+\phi\right) &=& 8 \pi \Sigma_T G a^2 \sum \rho_i \Delta_i \label{lensing_ft3}.
\end{eqnarray}

As usually done in the literature, assuming $i=m$ (only matter contributions and disregarding radiation and neutrinos), we have

\begin{eqnarray}
-\frac{k^2}{a^2}\left(\psi+\phi\right) = 8 \pi \Sigma_T G \rho_m \Delta_m \label{lensing_ft4},
\end{eqnarray}
where

\begin{eqnarray}
\label{sigma}
\Sigma_T = \frac{\mu_T}{2} \Big[ (1+R) + \frac{3 \Xi \mathcal{H}  \theta_m}{ \Delta_m} \Big].
\end{eqnarray}

This framework described above encodes the deviations from GR into two phenomenological functions, namely the effective gravitational coupling $\mu$ and the light deflection parameter $\Sigma$, which enter the Poisson and lensing equations, respectively. The function $\mu$ encodes the deviations of the gravitational interaction on the clustering of matter with respect to $\Lambda$CDM, while $\Sigma$ measures the deviation in the lensing gravitational potential. The $\Lambda$CDM model is recovered when $\mu = \Sigma = 1$. This methodology has been used to investigate efficiently the most diverse proposals of MG scenarios (see \cite{Roman_2013,Juan_2019,Ning_2020,Alex_2021,JAK_2021,Sakr_2022,FS_2020,NF_2022,Li_2018,LP_2016,Pace_2021,Noemi_2018} for a short list). 

To complete the set of linear perturbations, we also take into account the tensor perturbations modes \cite{Tomi_2018} 

\begin{equation}
\label{tensor}
h_{ij}^{\prime\prime}+2{\cal H}\left(1 + \frac{\Xi}{2a^2}\right)h_{ij}^{\prime}+k^2 h_{ij}=0.
\end{equation}

An important point is that the speed of GWs remains the same as in GR. Some direct consequences of GWs in $f(T)$ gravity have been investigated previously in \cite{LX_2018,SP_2018,TP_2022,ZZ_2021,Caso_2020,Saif_2021,Nunes_2019,Manuel_2018,Viktor_2018,SB_2021}.

In this work, we consider the power-law model given by  
\begin{equation}
f(T)= T + F(T)=T+\alpha(-T)^b,
\end{equation}
where $\alpha$ and $b$ are two model parameters \cite{Gabriel_2009}. This parametric form has been the most investigated in literature and perhaps the most viable model. Our main aim is not to exhaustively analyze different models and/or parameterizations, but present a new methodology and perceptions to analyze the $f(T)$ gravity framework. Therefore, we focus on the results of this specific model. Without loss of generality, the methodology can be applied to other parametric models of $f(T)$ gravity. Inserting this $f(T)$ form into Friedmann equation at present, we acquire a theoretical constraint on the parameter $\alpha$, viz.,

\begin{equation}
\alpha=(6H_0^2)^{1-b}\frac{\Omega_{F0}}{2b-1}.
\end{equation}

Thus, the full theory is completely specified from a single free parameter, i.e., the parameter $b$. In order to get the expansion rate of the Universe, i.e., the $H(z)$ function, we follow the same methodology as in \cite{SN_2013}.\\

We modify the \texttt{CLASS}~\cite{Blas:2011rf,Lesgourgues:2011re} code to introduce these perspectives for the $f(T)$ gravity model to be tested here with the observational data \footnote{We add the new features of the modified equations for background and perturbation  evolutions in \texttt{CLASS} by implementing the modifications in \texttt{background.c} and \texttt{perturbations.c}  modules, respectively. More specifically, the background equations (\ref{Mod1Ez})-(\ref{distortparam}), following the methodology of Ref. \cite{SN_2013} are implemented in \texttt{background.c}, and the Poisson and lensing equations (\ref{poisson})-(\ref{sigma}) and the single Einstein equation (\ref{tensor}) for tensor perturbations are implemented in the function \texttt{perturbations{\_}einstein}.}.

\section{Data and Methodology}
\label{data}

In order to derive constraints on the model baseline, we use the following datasets.
\begin{itemize}
\item \textbf{CMB}: From the \textit{Planck} 2018 legacy data release, we use the CMB measurements, viz., high-$\ell$ \texttt{Plik} TT likelihood (in the multipole range $30 \leq \ell \leq 2508$), TE and EE (in the multipole range $30 \leq \ell \leq 1996$),  low-$\ell$ TT-only ($2 \leq \ell \leq 29$), the low-$\ell$ EE-only ($2 \leq \ell \leq 29$) likelihood~\cite{Planck:2019nip}, in addition to the CMB lensing power spectrum measurements \cite{Planck:2018lbu}.

\item \textbf{BAO}: From the latest compilation of  Baryon Acoustic Oscillation (BAO) distance and expansion rate measurements from the SDSS collaboration, we use 14 BAO measurements, viz., the isotropic BAO measurements of $D_V(z)/r_d$ (where $D_V(z)$ and $r_d$ are the spherically averaged volume distance, and sound horizon at baryon drag, respectively) and anisotropic BAO measurements of $D_M(z)/r_d$ and $D_H(z)/r_d$ (where $D_M(z)$ and $D_H(z)=c/H(z)$ are the comoving angular diameter distance and  the Hubble distance, respectively), as compiled in Table 3 of \cite{eBOSS Collaboration}.

\item \textbf{BBN}: We use the state-of-the-art assumptions on Big
Bang Nucleosynthesis (BBN) data comprising of measurements of the primordial abundances of helium $Y_P$ \cite{Erik_2015} and the deuterium measurement $y_{DP} = 10^5 n_D/n_H$ \cite{Max_2018}. It is known that the BBN likelihood is sensitive to the constraints on the physical baryon density $\omega_b \equiv \Omega_bh^2$ and the effective number of neutrino species $N_{\rm eff}$, and we fix $N_{\rm eff}= 3.046$ in the present work. For theoretical predictions, the baseline likelihood uses the code \texttt{PArthENoPE 2.0} \cite{ParthENoPE_2018}.

\item \textbf{Pantheon}: We use the Type Ia Supernovae distance moduli measurements from the Pantheon sample~\cite{DMS_2018}, which constrain the uncalibrated luminosity distance $H_0d_L(z)$, or the slope of the late-time expansion rate (which in turn constrains $\Omega_m$).  
\end{itemize}
The theoretical apparent magnitude $m_B$ for a SN at redshift $z$ reads

\begin{eqnarray}
\label{distance_modulus}
m_B = 5 \log_{10} \left[ \frac{d_L(z)}{1 Mpc} \right] + 25 + M_B,
\end{eqnarray}
where $M_B$ is the absolute magnitude.

The calibrated Type Ia Supernova absolute magnitude $M_B$ is in general assumed to be truly a constant, i.e., the parameter $M_B$ should  be independent of redshift $z$. It has been argued that a possible variation of the absolute magnitude $M_B$ and equivalently of the absolute luminosity $L \sim 10^{-2M_B/5}$, could be due to a variation in the value of Newton’s constant $G$ \cite{Bravo_1,RSW_18}. This is due to the fact that the absolute luminosity is proportional to the Chandrasekhar mass $L \sim M_{\rm Chandra}$,  which depends on Newton’s constant $G$ as $L \sim G^{-3/2}$. Therefore, any modification of gravity will generate an 
effective gravitational constant in the form of $G_{\rm eff}$ that will induce a natural correction on the distance modulus. In the presence of a varying effective gravitational constant, the theoretical distance modulus defined by $\mu_{\rm{th}}(z) = m_B - M_B$, in view of eq.(\ref{distance_modulus}), is given by

\begin{equation}
\label{mb_ft}
 \mu_{\rm{th}}(z) =  5 \log_{10} d_L(z) + 25 + \frac{15}{4} \log_{10} \frac{G_{\rm eff}(z)}{G}.
\end{equation}

\begin{figure}[hbt!]
\begin{center}
\includegraphics[width=8.5cm]{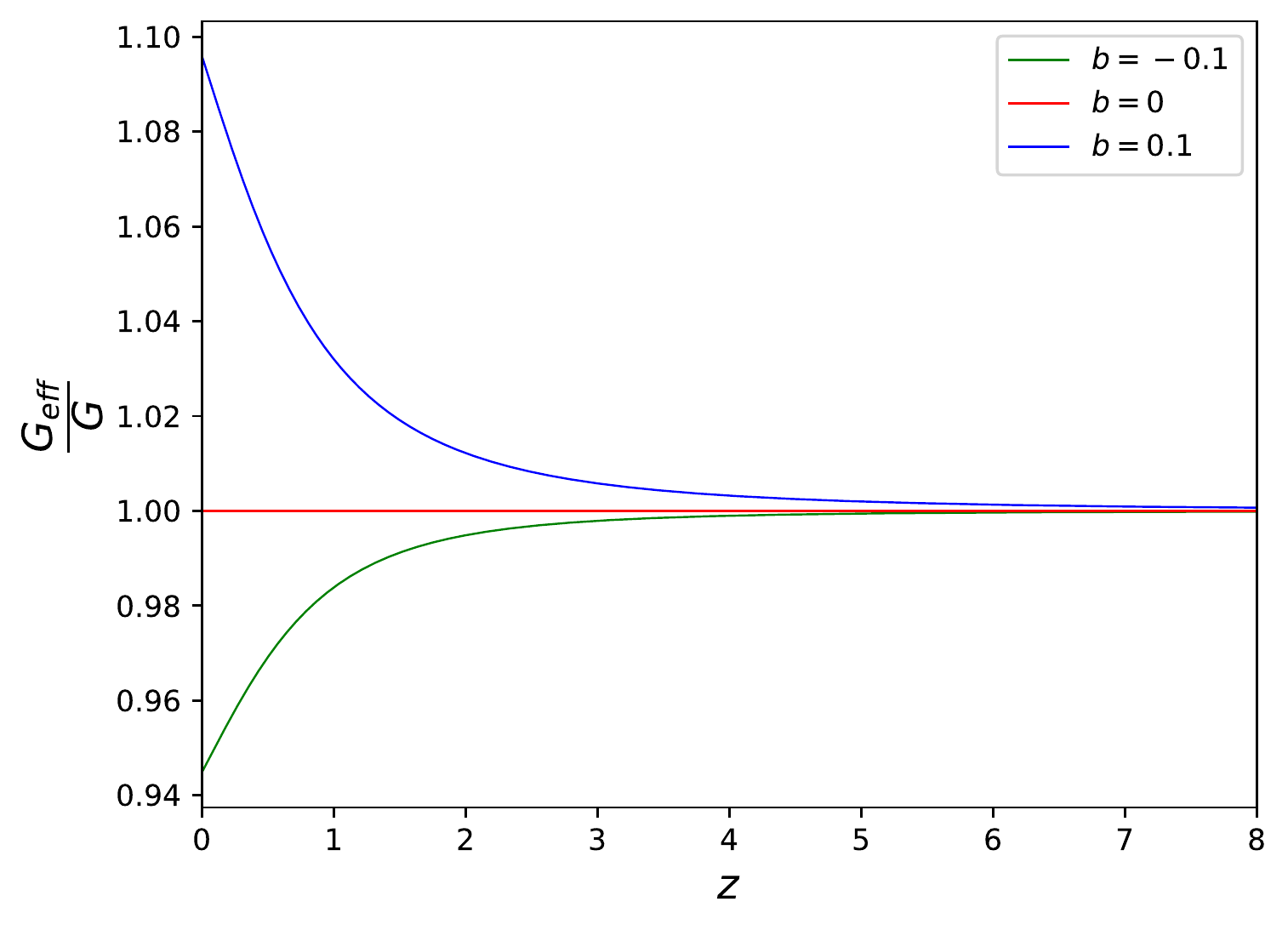} 
\caption{The effective gravitational coupling $\mu_T(z)=\dfrac{G_{\rm eff}(z)}{G}$ as a function of redshift $z$ for different and reasonable values of $b$. Other parameters are fixed to their Planck-CMB best fit values.} 
\label{fig:G_eff}
\end{center}
\end{figure}

Taking the quasi-static approximation and the modified Poisson equation in $f(T)$ gravity context, we have \cite{Rafael_2018,Wu_2012}

\begin{equation}
\frac{G_{\rm eff}(z)}{G} = \frac{1}{f_T} = \mu_T.
\end{equation}

Thus, it is possible that redshift dependence may carry useful information about the robustness of the determination of $H_0$ using the Pantheon sample and about possible modifications of $G_{\rm eff}(z)$ induced from the $f(T)$ gravity framework. 
Figure \ref{fig:G_eff} shows the behavior of the effective gravitational coupling as a function of $z$. As expected for very high $z$, we have $\mu_T = 1$, while at late times, deviations of gravitational force depend on the sign of the distortion parameter $b$, viz., gravity can be  weaker ($b < 0$), or stronger ($b > 0$) in comparison with GR for $z < 6$. The effect increases as $z$ decreases, as expected. Several authors have recently explored this possibility or similar approximations in the most diverse modified gravity scenarios \cite{Marra_2021,Skara_2021,Mario_2022,RPG_2021,Jeremy_2014,RSW_18,Daniel_2019,py_2022}. Here we explore these consequences in $f(T)$ gravity in the next section.

All cosmological observables are computed with \texttt{CLASS}~\cite{Blas:2011rf,Lesgourgues:2011re}. In order to derive bounds on the proposed scenarios, we modify the efficient and well-known cosmological package \texttt{MontePython}~\cite{Brinckmann:2018cvx}. In particular, we modify the Pantheon likelihood to incorporate the predictions given by eq. (\ref{mb_ft}). We assess the convergence of the MCMC chains using the Gelman-Rubin parameter $R-1$~\cite{Gelman:1992zz}, requiring $R-1<0.01$ for the chains to be converged.

For the analyses with the CMB data, the baseline reads:
\begin{eqnarray}
\label{baseline_CMB}
\{\omega_b, \omega_{\rm c}, \theta_s, A_s, n_s, \tau, b \}.
\end{eqnarray}
Here, $\omega_{\rm b}$ and $\omega_{\rm c}$ are the baryon and cold dark matter physical densities; $\theta_s$ is the angular acoustic scale; ${A_s}$ and $n_s$ are the amplitude and tilt of the initial curvature power spectrum at the pivot scale $k = 0.05$/Mpc; $\tau$ is the optical depth to reionization and $b$ is the free parameter of the $f(T)$ gravity model, quantifying the deviations from the $\Lambda$CDM model.

In the analyses without the CMB data, the baseline reads:
\begin{eqnarray}
\label{baseline_CMBw}
\{\omega_b, \omega_{\rm c}, b \}.
\end{eqnarray}

\begin{figure*}
\includegraphics[width=8cm]{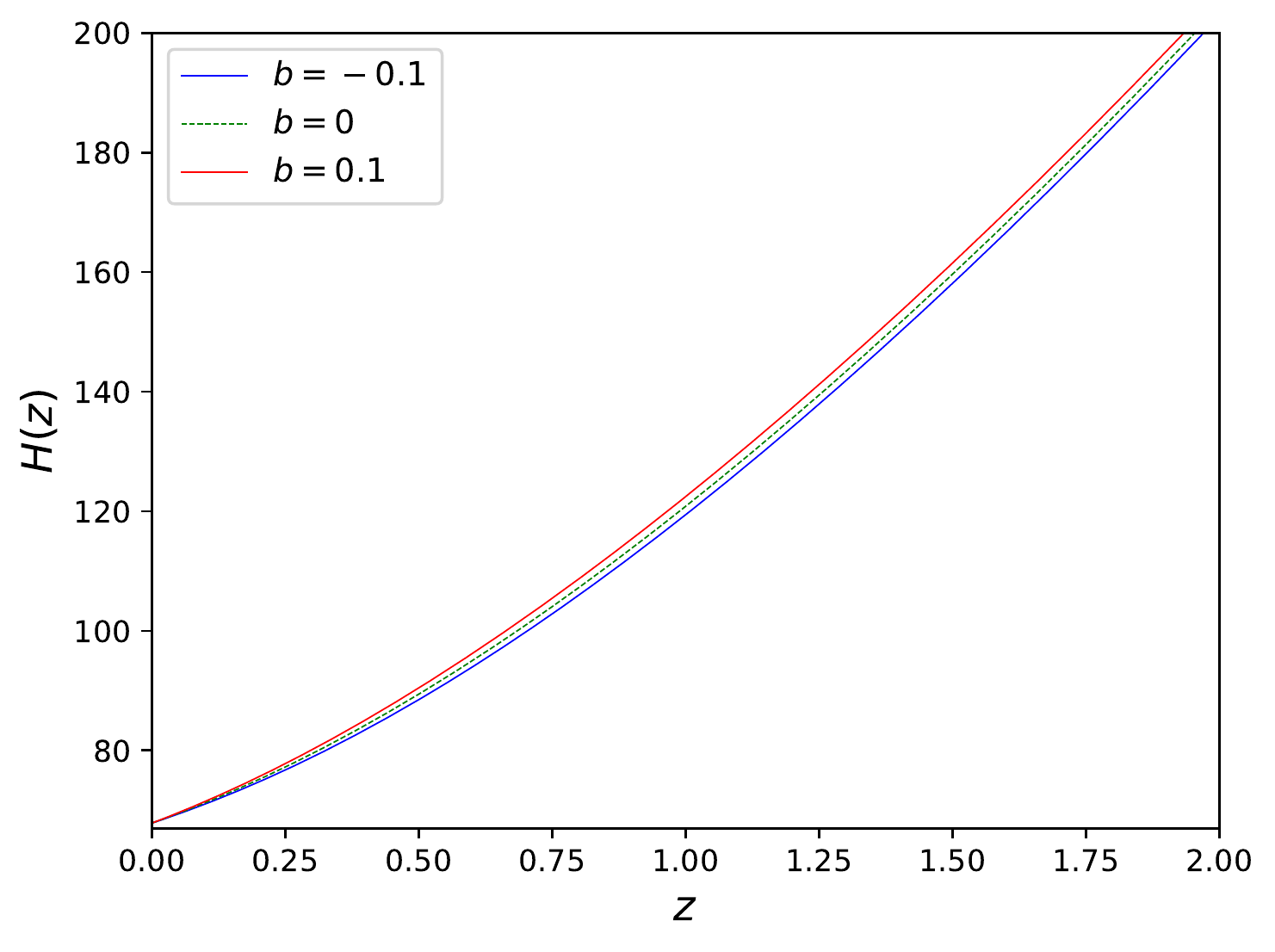} 
\includegraphics[width=8.5cm]{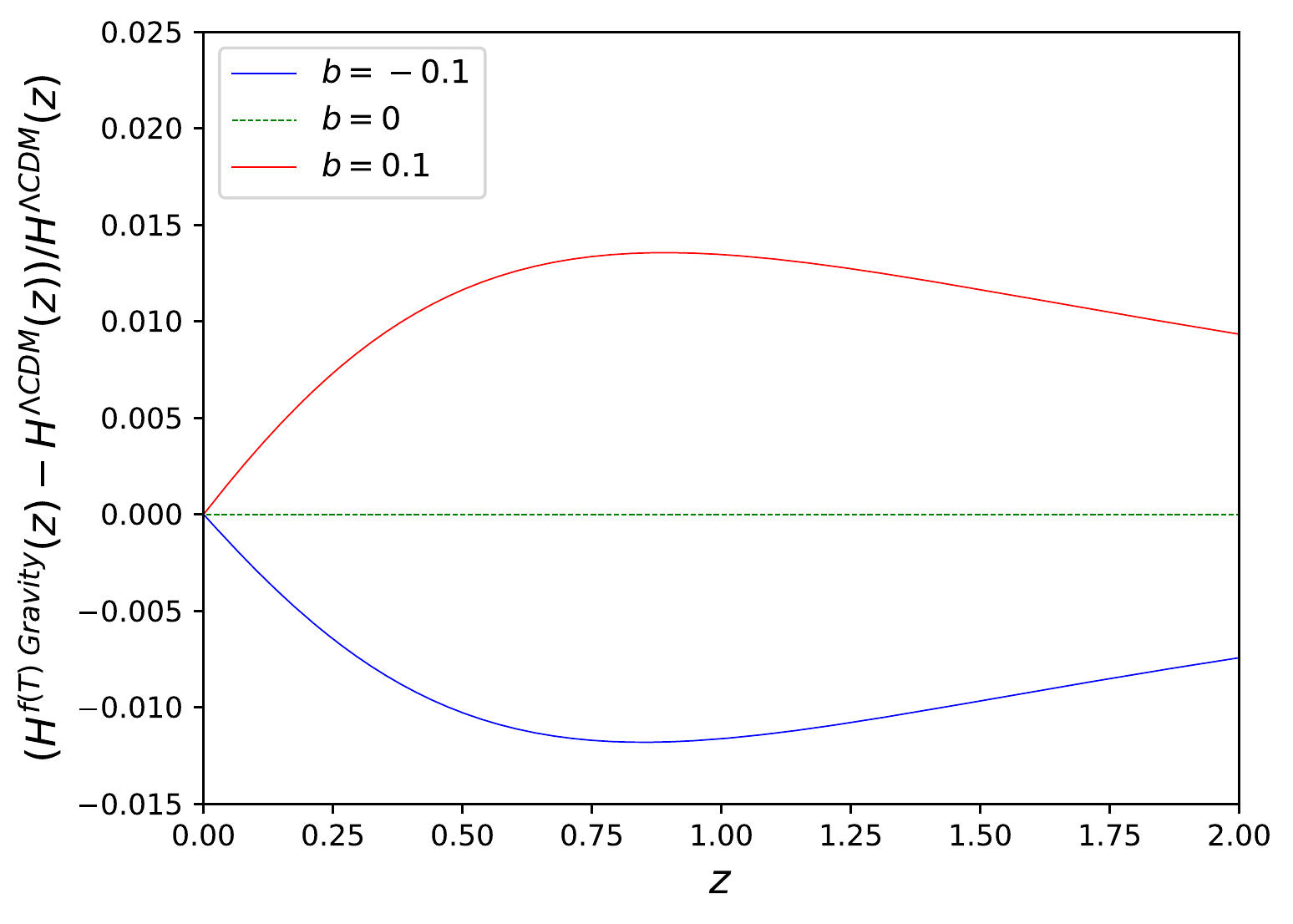}
\caption{Left panel shows the theoretical prediction for the expansion rate of the Universe at late times for reasonable and different values of $b$. The value $b=0$ corresponds to $\Lambda$CDM model.
Right panel shows the difference $(H^ {f(T)\, {\rm Gravity}}(z)-H^{\Lambda {\rm CDM}}(z))/H^{\Lambda {\rm CDM}}(z)$.}
\label{fig:1}
\end{figure*}

\begin{figure*}
\includegraphics[width=8cm]{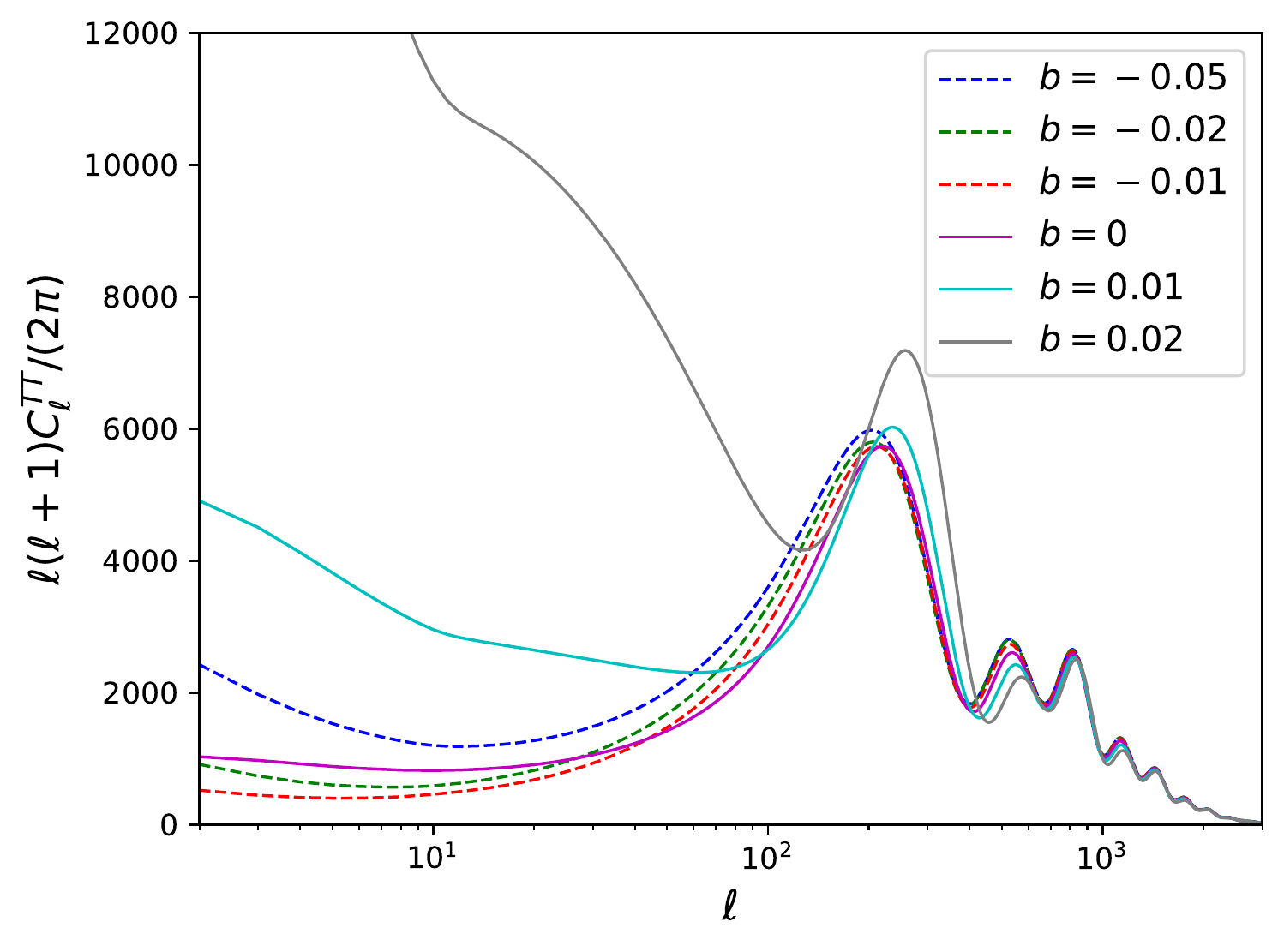}
\includegraphics[width=8cm]{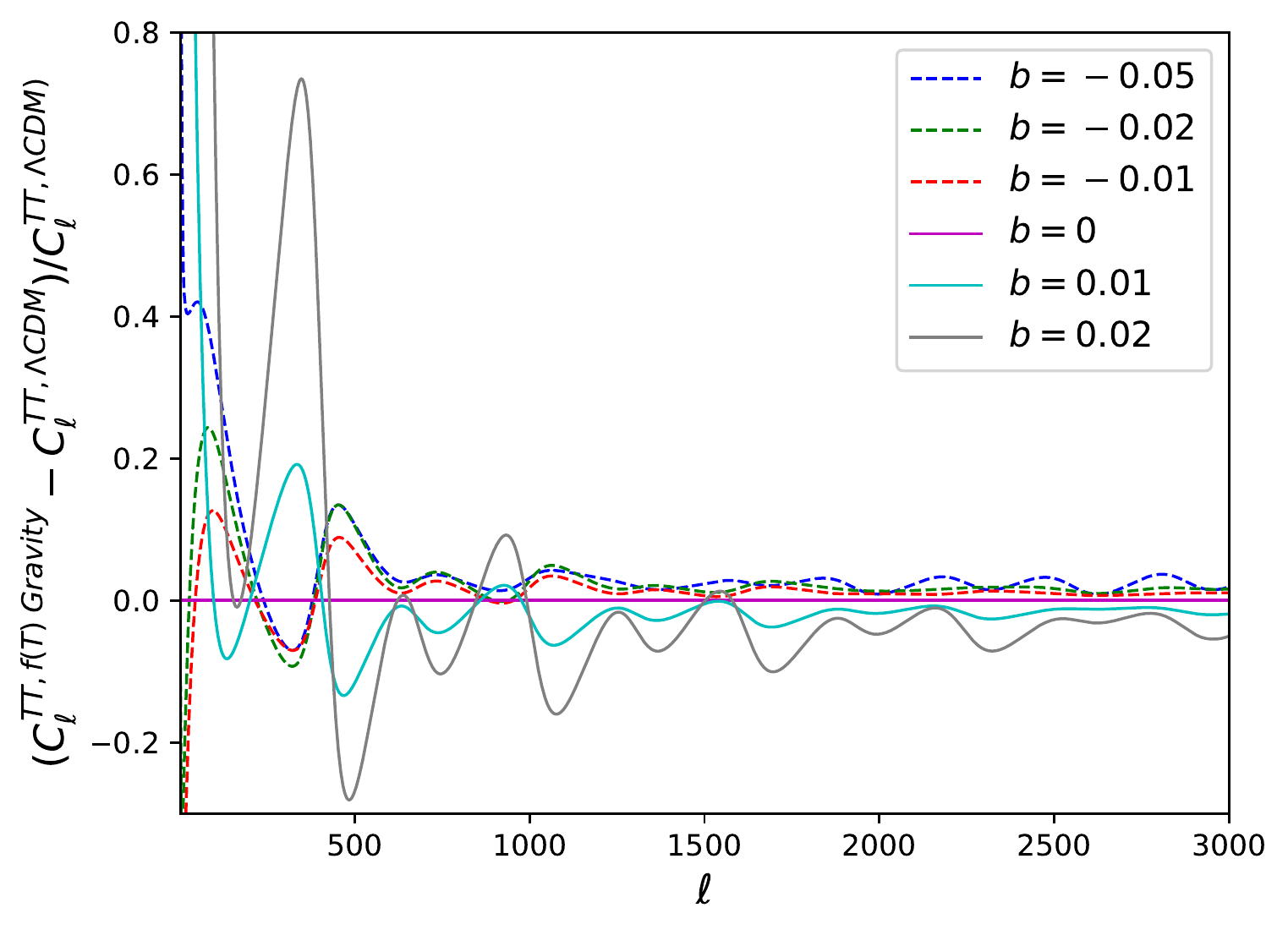}
\caption{Left panel shows the theoretical prediction for the CMB TT anisotropy power spectrum for different values of $b$. The value $b=0$ stands for $\Lambda$CDM model. Right panel shows the difference $(C^{{\rm TT}, f(T) \, {\rm Gravity}}_{\ell} - C^{\rm TT, \Lambda CDM}_{\ell})/C^{\rm TT, \Lambda CDM}_{\ell}$  for different values of $b$. }\label{fig:2}
\end{figure*}

\begin{figure*}
\includegraphics[width=8cm]{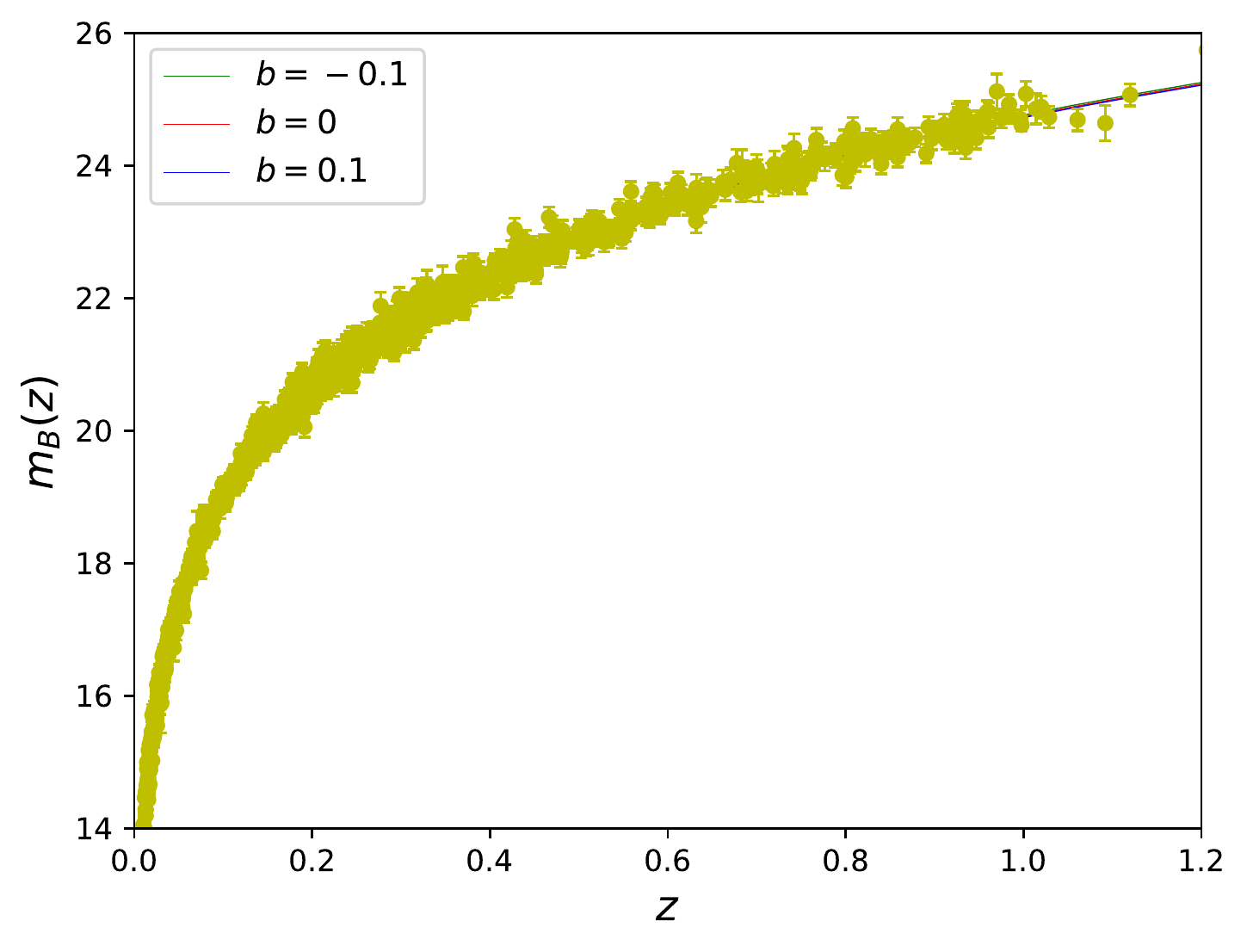}
\includegraphics[width=8.5cm]{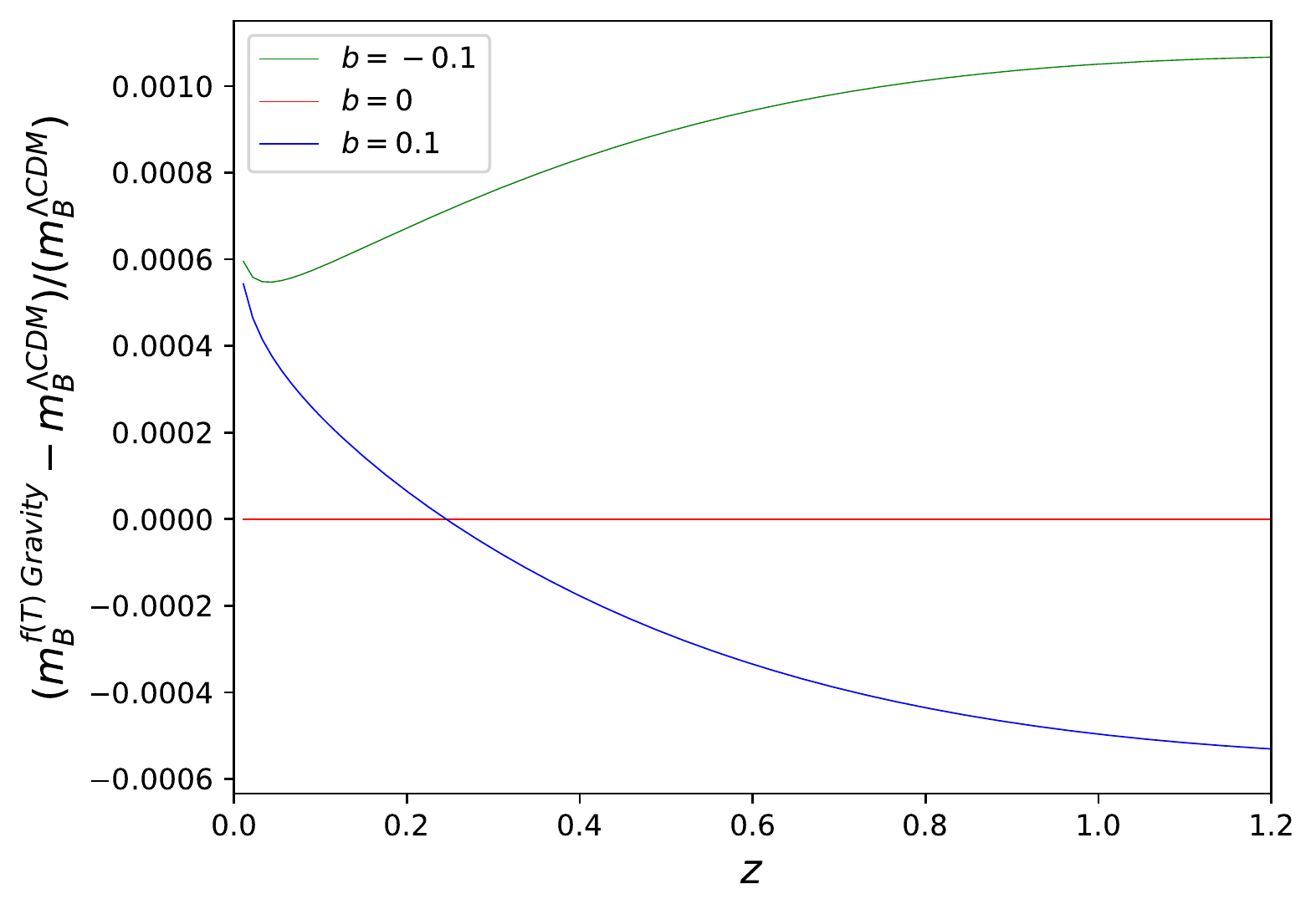}
\caption{Left panel shows the magnitude-redshift relation of the Pantheon SN sample in the redshift range $0 < z < 1.2$ for values of $b$ within the range of values summarized in Table \ref{table1}. Right panel shows the relative difference compared to $\Lambda$CDM model using the same values.}
\label{mb_data}
\end{figure*}

In our analyses, we assume all baseline parameters with wide ranges of flat priors. In what follows, we present and discuss our main results.
\begin{table*}[hbt!]
\centering
\caption{Constraints at 68\% CL on the free and some derived parameters in $f(T)$ gravity and standard $\Lambda$CDM models for BAO+BBN and BAO+BBN+Pantheon data. In the analyses with Panthon data, $M_B$ is the nuisance parameter. The parameter $H_{\rm 0}$ is measured in units of km s${}^{-1}$ Mpc${}^{-1}$. }\label{table1}
\scalebox{0.95}{
\begin{tabular}{|c|c|c|c|c|c|c|cc|cc}       
\hline
Model $\rightarrow$& \multicolumn{2}{|c|}{$\Lambda$CDM Model}  &\multicolumn{2}{|c|}{$f(T)$ Gravity Model}\\\hline
Parameter & BAO+BBN & BAO+BBN+Pantheon & BAO+BBN & BAO+BBN+Pantheon  \\ \hline
$10^{2}\omega_{\rm b }$ &  $2.234\pm 0.036$&$2.232\pm 0.036$&$2.234\pm 0.036$  & $2.231\pm 0.036 $  
     
    \\ 
    
$\omega_{\rm c }$  &$0.113^{+0.0100}_{-0.0130}$&$0.1114\pm 0.0089  $&$0.106^{+0.0100}_{-0.0130} $  & $0.109\pm 0.0100 $  
  
 \\

$b$  & ---& ---&$0.102^{+0.053}_{-0.060}$   & $0.044\pm 0.038 $
  
\\\hline

$M_{\rm B}$  &   --- & $-19.439\pm 0.037$ &---&$-19.431^{+0.028}_{-0.038}$
  
\\
\hline

$\Omega_{\rm{m} }$ & $0.296^{+0.017}_{-0.020}$&  $0.295\pm 0.014  $& $0.292^{+0.017}_{-0.021} $    & $0.290\pm 0.016 $ \\

$H_{\rm 0}$ & $67.5^{+1.1}_{-1.2}  $& $67.3\pm 1.0  $& $66.2^{+1.2}_{-1.4} $     & $66.8\pm 1.2 $  
 
\\

$G_{\rm{eff}}(z=0)$$/G$ & $1  $& $1    $& $1.117^{+0.044}_{-0.10}  $     & $1.040^{+0.033}_{-0.046}    $  
\\
\hline
\end{tabular}
}
\end{table*}

\begin{figure*}[hbt!]
\begin{center}
\includegraphics[width=12cm]{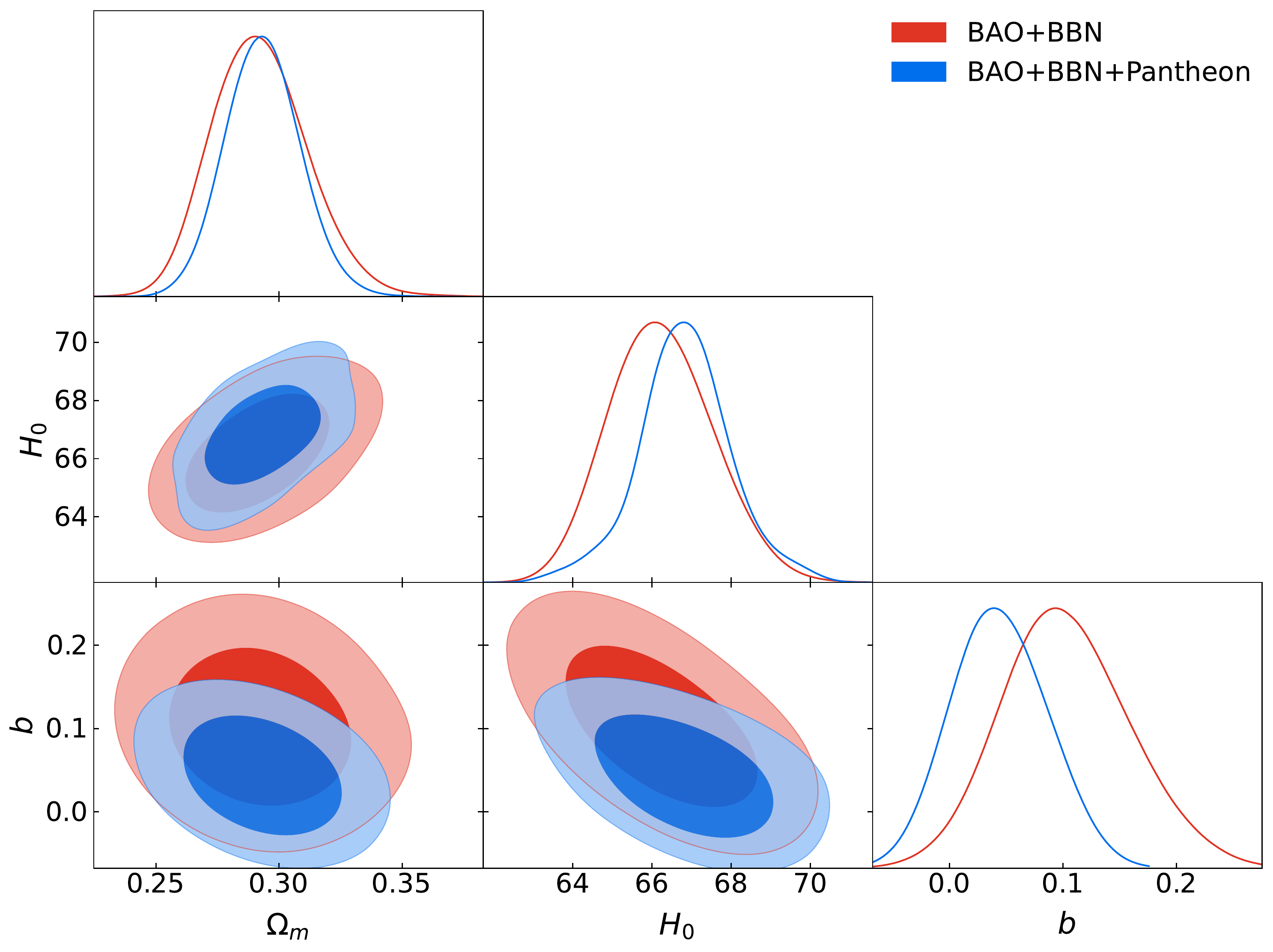} 
\caption{ One-dimensional and two-dimensional marginalized confidence regions (68\% CL and 95\% CL) for $b$, $\Omega_{\rm m}$ and $H_0$ 
obtained from the BAO+BBN and BAO+BBN+Pantheon data for the $f(T)$ gravity model. The parameter $H_0$ is in units of km s${}^{-1}$ Mpc${}^{-1}$.} 
\label{fig:BAO_SN}
\end{center}
\end{figure*}

\section{Main results and discussions}
\label{results}

Figure \ref{fig:1} on the left panel shows the theoretical prediction for the expansion rate of the Universe at late times for reasonable and different values of $b$. On the right panel, we show a relative difference $(H^ {f(T)\;\rm Gravity}(z)-H^{\rm \Lambda CDM}(z))/H^{\rm \Lambda CDM}(z)$. We can note that for $b > 0$ ($< 0$), the contributions due to the effective dark energy terms induced for the torsional gravitational modification corrections  will make the Universe expand faster (slower), respectively. For a correction of $\mathcal{O}(0.1)$, we expect a difference of $\sim$1.5\% compared to $\Lambda$CDM model.

Figure \ref{fig:2} on the left panel shows the theoretical prediction for the CMB TT anisotropy power spectrum for different values of $b$. While drawing this plot, we fix all common parameters baseline to Planck-$\Lambda$CDM values, and then adopt different values of $b$. The $f(T)$ gravity under consideration here will affect CMB power spectrum in three ways. First, the change in the time evolution of metric potentials $\phi$ and $\psi$ will modify the early integrated Sachs–Wolfe effect. For modes inside the  horizon, these early time contributions tend to decrease as a function of $l$. The first peak localization can be significantly affected by this effect. Thus, different values of $b$ (in magnitude and sign) can differently change  the position of the first acoustic peak. The peak location will be changed by the angular diameter distance at decoupling, which depends on the expansion history after decoupling. Consequently, the scenario studied here will be changed due to the correction in the $H(z)$ function at late times. 

\begin{table*}[hbt!]
\centering
\caption{Constraints at 68\% CL on  the free and some derived parameters of the $f(T)$ gravity and standard $\Lambda$CDM models from CMB, CMB+BAO and CMB+BAO+Pantheon data. In the analyses with Panthon data, $M_B$ is the nuisance parameter. The parameter $H_{\rm 0}$ is measured in units of km s${}^{-1}$ Mpc${}^{-1}$.}\label{table2}
\scalebox{0.91}{
\begin{tabular}{|c|c|c|c|c|c|c|cc|cc}       
\hline
Model $\rightarrow$& \multicolumn{3}{|c|}{$\Lambda$CDM Model}  &\multicolumn{3}{|c|}{$f(T)$ Gravity Model}\\\hline
Parameter & CMB & CMB+BAO & CMB+BAO+Pantheon & CMB & CMB+BAO & CMB+BAO+Pantheon  \\ \hline
$10^{2}\omega_{\rm b }$ & $2.236\pm 0.015$&  $2.244\pm 0.013 $ &$2.242\pm 0.013 $&$2.239\pm 0.014$  & $2.238^{+0.013}_{-0.014}$ &  
     $2.241\pm 0.013$        \\
    
$\omega_{\rm c }$  &$0.1200\pm 0.0012$&$0.1193\pm 0.0010 $& $0.11934\pm 0.00093$  &$0.1196\pm 0.0016 $   &$0.1193\pm 0.0001 $ 
 & $0.11894^{+0.00086}_{-0.00100}$ 
 \\ 
 
$100 \theta_{s } $  & $1.04186\pm 0.00030 $ &$1.04190\pm 0.00029 $& $1.04194\pm 0.00028$ &$1.04200\pm 0.00039$     &$1.04195\pm 0.00032$   
&    $1.04212\pm 0.00035$ 
\\

$\ln(10^{10}A_{\rm s})$ &   $3.045^{+0.013}_{-0.016}$& $3.048\pm 0.015 $ &$3.047\pm 0.015  $ & $3.045\pm 0.014$   & $3.046^{+0.013}_{-0.016}$ 
&$3.045\pm 0.014$  
\\

$n_{s } $  & $0.9653\pm 0.0042  $&$0.9669^{+0.0039}_{-0.0035}$&$0.9665\pm 0.0037$&$0.9667\pm 0.0053$     & $0.9669^{+0.0046}_{-0.0042}$ 
&     $0.9688\pm 0.0039$  
\\

$\tau_{\rm reio } $  &  $0.0549^{+0.0067}_{-0.0082}$&  $0.0563\pm 0.0077$&  $0.0561\pm 0.0076$& $0.0544^{+0.0070}_{-0.0078}$   & $0.0560^{+0.0062}_{-0.0080}$ 
&$0.0557^{+0.0066}_{-0.0074}$   
\\

$b$  & ---& ---&  ---&$\left(0.8^{+2.4}_{-2.1}\right)\times{10^{-4}}$   & $\left(0.8^{+2.0}_{-2.0}\right)\times{10^{-4}}$  
&$\left(1.7^{+2.0}_{-2.0}\right)\times{10^{-4}}$  
\\\hline

$M_{\rm B}$  &   --- & --- & $-19.418\pm 0.012$ & ---&---
& $-19.413^{+0.013}_{-0.012} $    
\\

\hline
 
$\Omega_{\rm{m} }$ & $0.3151\pm 0.0074$&$0.3109\pm 0.0060 $ & $0.3109\pm 0.0057$& $0.3123\pm 0.0098 $     & $0.3107\pm 0.0060$   & $0.3082^{+0.0052}_{-0.0065}$ \\

$H_{\rm 0}$ &  $67.38\pm 0.54 $& $67.69\pm 0.44 $&$67.69^{+0.38}_{-0.43}$& $67.60\pm 0.71 $      & $67.69\pm 0.44 $ 
& $67.88^{+0.47}_{-0.41}$    
\\

$G_{\rm{eff}}(z=0)$$/G$ &  $1  $& $1$&$1$& $1.00006^{+0.00016}_{-0.00015} $      & $1.00006\pm 0.00014 $ 
& $1.00012\pm 0.00014 $  \\

\hline                                                
\end{tabular}
}

\end{table*}

\begin{figure*}[hbt!]
\begin{center}
\includegraphics[width=15cm]{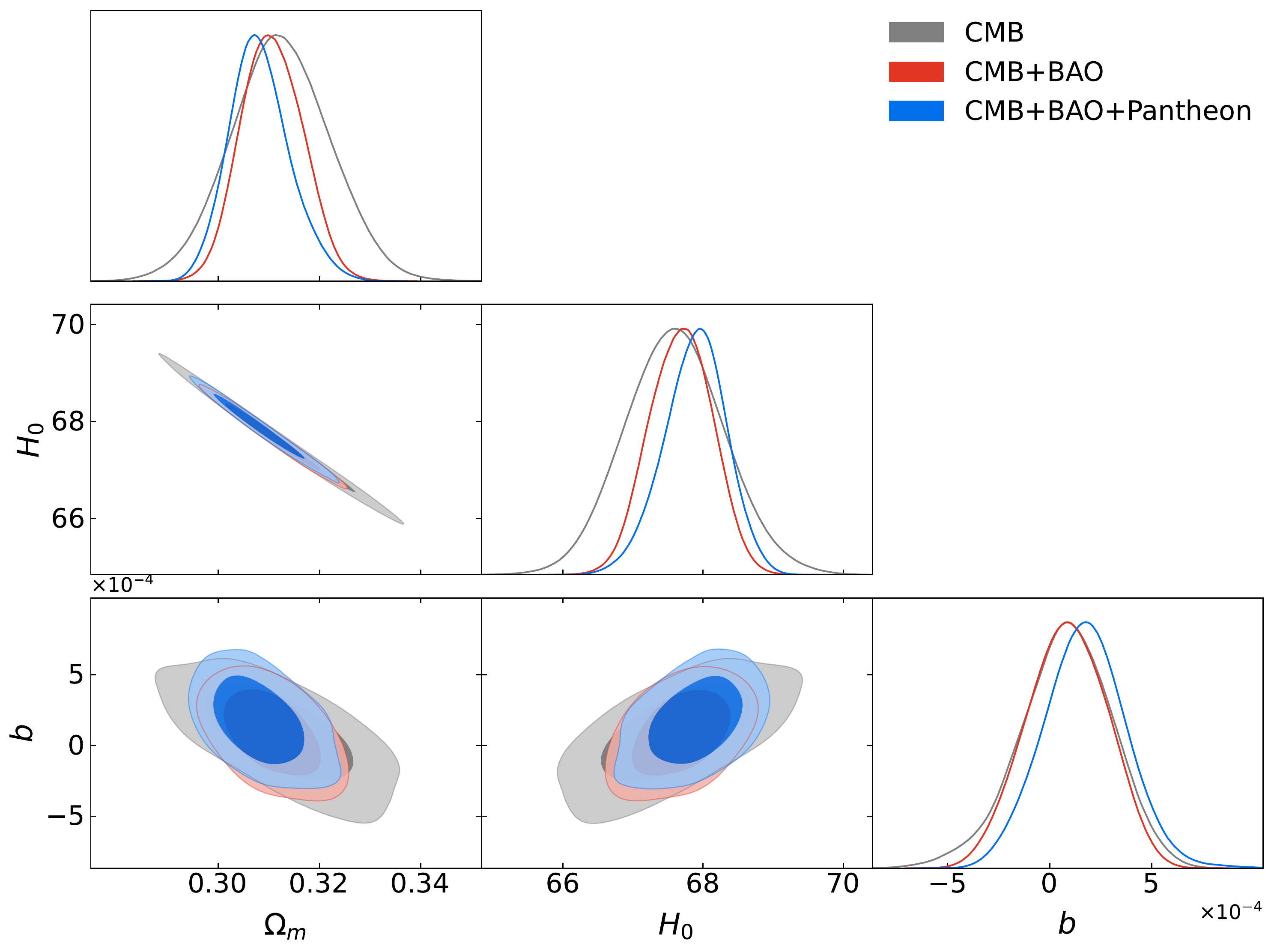} 
\caption{One-dimensional and two-dimensional marginalized confidence regions (68\% and 95\% CL) for $b$, $\Omega_m$ and $H_0$ 
obtained from the CMB, CMB+BAO, and CMB+BAO+Pantheon data for the $f(T)$ gravity model.  The parameter $H_0$ is in units of km s${}^{-1}$ Mpc${}^{-1}$.}
\label{fig:CMB+}
\end{center}
\end{figure*}

Secondly, it is important to note that the residual time variation of $\phi$ and $\psi$ at late times, when the effective dark energy changes the Universe’s expansion-rate behaviour, the metric fluctuations start decaying again. Thus, the model will also change the late integrated Sachs–Wolfe contribution. This late effect is present on all scales, because the effective dark energy domination produces a decay of metric fluctuations at all wavelengths, but with peaks at $l = 2$, and predominant effects on very large scales, i.e., $l < 40$. This effect is also attenuated due to changes in $H(z)$ function at late times. Thus, the model tends to deviate significantly from the $\Lambda$CDM at large scales.  The predominance of these modifications at large scales is clear in Figure \ref{fig:2}. Such effects on CMB power spectrum are also quantified in \cite{Benetti_2020}. Therefore, even small corrections of the order of $b \sim$ $\mathcal{O}(0.01)$ can significantly affect CMB power spectrum. Thus, the observational constraints from CMB data must be predominant. Third main effect comes from the tensor perturbations, which are known to contribute to temperature anisotropies on large angular scales and to E-type and B-type polarization anisotropies on all scales. For reasonable values of the free parameter $b$ of our model, these effects must be negligible on the temperature and polarization anisotropies that we are going to use to constrain the model, i.e., the Planck-CMB data. We keep the predictions of the tensor modes simply for consistency. Important aspects in this regard are investigated previously in \cite{SP_2018}. On the right panel in Figure \ref{fig:2}, we show the relative difference $(C^{\rm TT, f(T) \, Gravity}_{\ell} - C^{\rm TT, \Lambda CDM}_{\ell})/C^{\rm TT, \Lambda CDM}_{\ell}$. We notice that the parameter $b$ can affect the CMB temperature considerably.

Table \ref{table1} reports a summary of the analyses considering the BAO+BBN and BAO+BBN +Pantheon data combinations. We also show the constraints on the $\Lambda$CDM model for comparison. The BBN information in combination with BAO is considered to break any possible degeneracy in the $H_0-\Omega_{\rm m}$ plane
because we know that BAO+BBN can constrain these parameters very well. Thus we first analyze BAO+BBN combination, i.e., without taking into account the new perspectives on the Type Ia Supernovae distance moduli. We note that the parameter $b$ can be non-null at 1$\sigma$ CL, but fully compatible with $\Lambda$CDM model for greater statistical significance. With the addition of Pantheon data, we can note improvement on the full baseline, and in particular the tendency of $b$ to the null values. This is due to the introduction of new factors in eq.\eqref{mb_ft}. Figure \ref{mb_data} on the left panel shows the magnitude-redshift relation of the Pantheon SN sample in the redshift range $0 < z < 1.2$ for values of $b$ within the range of best-fit values summarized in table \ref{table1}. On the right panel, we show the relative difference from the $\Lambda$CDM model using the same values. Figure \ref{fig:BAO_SN} shows the parametric space at 68\% CL and 95\% CL  for the parameters of interest of the model. The constraints (68\% and 95\% CLs) on the free parameter of the theory from the joint analysis BAO+BBN+Pantheon read $b = 0.044 ^{+0.038+0.069}_{-0.038-0.066}$. One may expect that the calibrated absolute magnitude $M_{\rm B}$ could undergo some modification. However, we do not find significant changes on $M_{\rm B}$ (and consequently also on $H_0$). Thus, in the light of these new perspectives, the model under consideration here does not alleviate the tension in $M_{\rm B}$ and/or $H_0$.

Table \ref{table2} summarizes the constraints at 68\% CL on  the free and some derived parameters of the $f(T)$ gravity and standard $\Lambda$CDM models from CMB, CMB+BAO and CMB+BAO+Pantheon data. Figure \ref{fig:CMB+} shows the parametric space at 68\% CL and 95\% CL  for some parameters of interest. As previously mentioned, any variation in $b$ can induce significant changes in the CMB spectrum. Looking at the percentage of changes that can occur compared to $\Lambda$CDM model, quantified in Figure \ref{fig:2}, we expect $b << 1$. In all the analyses with the CMB data from Planck 2018 release, here we find $b \sim$ $\mathcal{O}(10^{-4})$, and a complete statistical agreement of the $f(T)$ gravity model with the $\Lambda$CDM. We see that the full baseline does not show significant changes compared to $\Lambda$CDM, and notice only small increments on the error bars due to the additional free parameter $b$. The joint analyses with the BAO and Pantheon data  provide tight constraints on the parameters of the models.  The final constraints (68\% and 95\% CLs) on the free parameter $b$ of the theory from the full joint analysis CMB+BAO + Pantheon read $b = \left(1.7^{+2.0+4.0}_{-2.0-3.9}\right)\times{10^{-4}}$. From Figure \ref{fig:CMB+}, we notice a positive correlation of $b$ with $H_0$. So larger values of $b$ correspond to the larger values of $H_0$. In our present analyses, the parameter $b$ is left free with flat priors, and we find $b \sim$ $\mathcal{O}(10^{-4})$, the most robust constraint on $b$ in the context of $f(T)$ gravity scenerio compared to earlier studies on $f(T)$  gravity with different perspectives \cite{Benetti_2020,Aliya_2022,David_2020,Amr_2021,RCN_2020,Mota_2020,Cai_2020,SB_2018,Zant_2018,Santos_2021,Nashed_2019,Briffa_2022,ZZ_2021,FK,SC,RC,rf_17,QI,Fotios_2019,Bing,Hashim_2021,Rafael_2018,SN_2013}. Finally, therefore, in light of these new perspectives, the $f(T)$ gravity scenario investigated in the present work is practically indistinguishable from the $\Lambda$CDM model.

It may be noted that the power-law model of $f(T)$ gravity considered here is studied with the full CMB data (Planck 2015 release) and BAO data in \cite{Rafael_2018}. In the context of our present study, it is important to highlight the recent work \cite{Benetti_2020}, where the background expansion history of the power-law model was investigated using Planck-CMB 2018 and its combination with several other data-sets. Now, in the study presented here, we have studied this model with full CMB (Planck 2018 release) data and an updated compilation of BAO data. We have also taken into account the new perspectives in the light of the SN sample, while correcting the theoretical predictions of the distance moduli. In addition, we have also considered some new perspectives at the level of linear perturbations not before exploited. Therefore, the theoretical methodology in combination with these new/updated data samples are studied for the first time in this work in the context of $f(T)$ gravity. Some differences of our main motivations and results can be pointed out in relation to previous works: i) We have taken into account the evolution of the scalar and tensor perturbations modes, leading to more robust/restrictive constraints on $b$, when the model is confronted with CMB data; ii) Our current BAO sample is vastly robust compared to all other previous works. iii) Our SN sample is corrected by taking into account the evolution of $G_{\rm eff}$ effects, leading to more robust constraints on the parameter $b$. In \cite{Benetti_2020}, the authors robustly constrain the GR distortion parameter to $b \sim$ $\mathcal{O}(10^{-2})$ while in our present study, we get the most tight constraint  $b\sim$ $\mathcal{O}(10^{-4})$ ever obtained in the context of $f(T)$ gravity power-law model. Further, a strong degeneracy between $H_0$ and $b$, and the use of $H_0$ prior (R19) lead to higher values of $H_0$ in \cite{Benetti_2020}. In our present analyses, we have not used Dark Energy Survey (DES) data as the galaxy clustering and weak gravitational sample should be used to test modified gravity only after  modeling the matter power spectrum to incorporate the modified gravity effects correctly, mainly at non-linear level predictions.

One can apply the new methodology presented here on various $f(T)$ gravity models to obtain more robust constraints on the distortion parameter $b$. For instance, here we have obtained very robust constraints on the distortion parameter $b$ of power-law $f(T)$ gravity model compared to previous studies.

\section{Final Remarks}

The increasing quantity and quality of available data allows us to test GR at cosmological scales with good accuracy. Several extensions of GR have been proposed as alternatives to the standard cosmological model despite of the $\Lambda$CDM model successes on several fronts. In the present work, we have investigated the effects of scalar linear perturbations parameterized in terms of the functions $\Sigma$ and $\mu$ in the context of modified teleparallel gravity on the CMB anisotropies power spectrum. Also, we have carried out our analyses by considering $G_{\rm eff}$ function in the SN distance modulus relation through the Pantheon sample of type Ia Supernovae. Both aspects are investigated in the present work for the first time. We have noticed that a very small correction in $f(T)$ gravity, beyond the $\Lambda$CDM model, is able to generate the fluctuations of CMB temperature. But from the analyses of $f(T)$ gravity power-law scenario with the full Planck-CMB 2018 data, we have found that these corrections are practically indistinguishable from the $\Lambda$CDM prediction. Our new methodology with the updated and new data sets has offered a more robsut constraint on the power-law model of $f(T)$ gravity compared to the previous study \cite{Rafael_2018}. We have obtained the strongest limits ever reported on $f(T)$ gravity power-law scenario at the cosmological level.  Finally, the strong constraint on $b$ from CMB data is not the final conclusion on $f(T)$ gravity, in general. In fact, we have introduced a new cosmological methodology never before explored in the context of $f(T)$ gravity, using full CMB and SN data, and presented the analysis of the most popular and used model, namely the power-law model from the literature of $f(T)$ gravity. Following our new observational methodology of $f(T)$ gravity, one may look for testing various $f(T)$ gravity models, especially in the context of cosmological tensions (e.g. \cite{Hashim_2021}).\\

\label{final}

\acknowledgments
The authors thank E.O. Colgain for useful discussions. S.K. gratefully acknowledges support from the Science and Engineering Research Board (SERB), Govt. of India (File No.~CRG/2021/004658). R.C.N thanks the CNPq for partial financial support under the project No. 304306/2022-3. P.Y. is supported by Junior Research Fellowship (CSIR/UGC Ref. No. 191620128350) from University Grant Commission, Govt. of India.

\end{document}